\documentclass[aps,prl,twocolumn,10pt,amsmath,amssymb,superscriptaddress,nofootinbib]{revtex4-1}
\usepackage{graphicx}
\usepackage{dcolumn}
\usepackage{bbm}
\usepackage{amsmath,amsfonts,amssymb}
\usepackage[justification=justified]{caption}
\usepackage{wrapfig}
\usepackage{svg}
\usepackage[colorinlistoftodos]{todonotes}
\usepackage[colorlinks,linkcolor=black, citecolor=black,urlcolor=black,bookmarks=false,hypertexnames=true]{hyperref} 
\usepackage{color}

\begin{document}

%comment out the line below and uncomment the following one to toggle blue changes on/off
\newcommand{\dggedits}[1]{\textcolor{blue}{#1}}
%\newcommand{\dggedits}[1]{#1}

% Dirac notation
\newcommand{\ket}[1]{\ensuremath{|#1\rangle\mkern-1mu}}
\newcommand{\bra}[1]{\ensuremath{\mkern-1mu\langle#1|}}
\newcommand{\ketbra}[2]{\ensuremath{|#1\rangle\mkern-5mu\langle#2|}}

%\title{Extraction of molecular excited states and error mitigation via the variational quantum eigensolver and the Quantum Subspace Expansion}
\title{Robust determination of molecular spectra on a quantum processor}
%Robust calculation of ground and excited state molecular energies via variational quantum simulation

\author{J. I. Colless}
	\affiliation{Quantum Nanoelectronics Laboratory, Department of Physics,
University of California, Berkeley CA 94720, USA.}
	\affiliation{Center for Quantum Coherent Science, University of California, Berkeley CA 94720, USA.}
\author{V. V. Ramasesh}
	\affiliation{Quantum Nanoelectronics Laboratory, Department of Physics,
University of California, Berkeley CA 94720, USA.}
	\affiliation{Center for Quantum Coherent Science, University of California, Berkeley CA 94720, USA.}
\author{D. Dahlen}
	\affiliation{Quantum Nanoelectronics Laboratory, Department of Physics,
University of California, Berkeley CA 94720, USA.}
	\affiliation{Center for Quantum Coherent Science, University of California, Berkeley CA 94720, USA.}
\author{M. S. Blok}
	\affiliation{Quantum Nanoelectronics Laboratory, Department of Physics,
University of California, Berkeley CA 94720, USA.}
	\affiliation{Center for Quantum Coherent Science, University of California, Berkeley CA 94720, USA.}
\author{J. R. McClean}
	\affiliation{Computational Research Division, Lawrence Berkeley National Laboratory, Berkeley, CA 94720, USA.}
\author{J. Carter}
	\affiliation{Computational Research Division, Lawrence Berkeley National Laboratory, Berkeley, CA 94720, USA.}
\author{W. A. de Jong}
	\affiliation{Computational Research Division, Lawrence Berkeley National Laboratory, Berkeley, CA 94720, USA.}    
\author{I. Siddiqi}
\email[To whom correspondence should be addressed: ]{\qquad\qquad irfan\_siddiqi@berkeley.edu}
	\affiliation{Quantum Nanoelectronics Laboratory, Department of Physics,
University of California, Berkeley CA 94720, USA.}
	\affiliation{Center for Quantum Coherent Science, University of California, Berkeley CA 94720, USA.}
    	\affiliation{Materials Science Division, Lawrence Berkeley National Laboratory, Berkeley, CA 94720, USA.}

\begin{abstract}

Harnessing the full power of nascent quantum processors requires the efficient management of a limited number of quantum bits with finite lifetime. Hybrid algorithms leveraging classical resources have demonstrated promising initial results in the efficient calculation of Hamiltonian ground states --- an important eigenvalue problem in the physical sciences that is often classically intractable. In these protocols, a Hamiltonian is parsed and evaluated term-wise with a shallow quantum circuit, and the resulting energy minimized using classical resources. This reduces the number of consecutive logical operations that must be performed on the quantum hardware before the onset of decoherence. We demonstrate a complete implementation of the Variational Quantum Eigensolver (VQE), augmented with a novel Quantum Subspace Expansion, to calculate the complete energy spectrum of the H$_2$ molecule with near chemical accuracy. The QSE also enables the mitigation of incoherent errors, potentially allowing the implementation of larger-scale algorithms without complex quantum error correction techniques.

\end{abstract}

\maketitle

Quantum computing, the field of physics dedicated to harnessing quantum phenomena to process information, is rapidly progressing along the path from theoretical curiosity to practical technology.  Recent experimental progress has been swift, with successful demonstrations of proof-of-concept algorithms on a range of fledgling quantum processors comprised of natural or engineered quantum spins~\cite{Monz, Debnath, Cai, Lucero, Takita}. However, even in leading architectures such as superconducting circuits and trapped ions, state-of-the-art systems comprise only few to tens of qubits---the quantum analog of classical bits---and are difficult to control with high precision.  For gate-based quantum processors to be competitive with, or outperform their classical counterparts, both qubit number and gate fidelity must increase significantly~\cite{Wecker, Fowler}.  Indeed, much of the field is currently focused on the design of a multi-qubit architecture capable of demonstrating an unambiguous quantum speedup over classical computers.

Recent theoretical advances suggest that a hybrid approach---judiciously dividing a computation between quantum and classical resources---will likely find utility in specific applications prior to the emergence of universal quantum computation~\cite{McClean, Li,Bauer,Tran}. 
One such example is calculating the ground-state energy of complex chemical systems, such as is often required in photovoltaics, biological reactions, and catalyst design.  Based on the quantum variational principle---that the ground-state wavefunction of any Hamiltonian minimizes the expected energy~\cite{Griffiths}---an iterative protocol, known as the Variational Quantum Eigensolver (VQE) was developed. This approach uses a classical optimization routine to minimize the expected energy of candidate wavefunctions, using the quantum hardware to evaluate the expected energy. Essentially the VQE leverages the unique capacity of even shallow quantum circuits to prepare entangled states from which efficient classical sampling is not known to be possible.  

Essential ingredients of the VQE algorithm have recently been demonstrated on a variety of experimental platforms~\cite{Peruzzo, Du, Shen, Wang, Omalley, Santagati, Kandala}. These initial experiments indicate a robustness to systematic control errors (so-called coherent errors) which would preclude fully quantum calculations, as well as a manageable scaling of quantum circuit depth with Hamiltonian complexity~\cite{Santagati, Kandala}. However, work to date has focused primarily on calculating molecular ground-state energies; while extending the VQE approach to find excited states has been demonstrated in the optical domain, it required additional qubits, complicated multi-qubit control, and additional variational searches~\cite{Santagati}. Furthermore, while a theoretical scheme for mitigating stochastic, incoherent errors has been proposed, this has yet to be verified experimentally~\cite{JarrodMollie}. In this work we demonstrate this extension of the VQE algorithm using a quantum processor comprising two superconducting transmon qubits, with real-time classical optimization augmented by a novel quantum subspace expansion (QSE) protocol. We diagonalize the electronic structure Hamiltonian of the hydrogen molecule over a wide range of nuclear separations and demonstrate the ability of this extended algorithm to calculate excited-state energies and partially mitigate stochastic error channels, attaining near chemical accuracy ($1.6\times10^{-3}$ H) in the calculated energy spectrum.

\begin{figure*}
  \includegraphics[width=\textwidth]{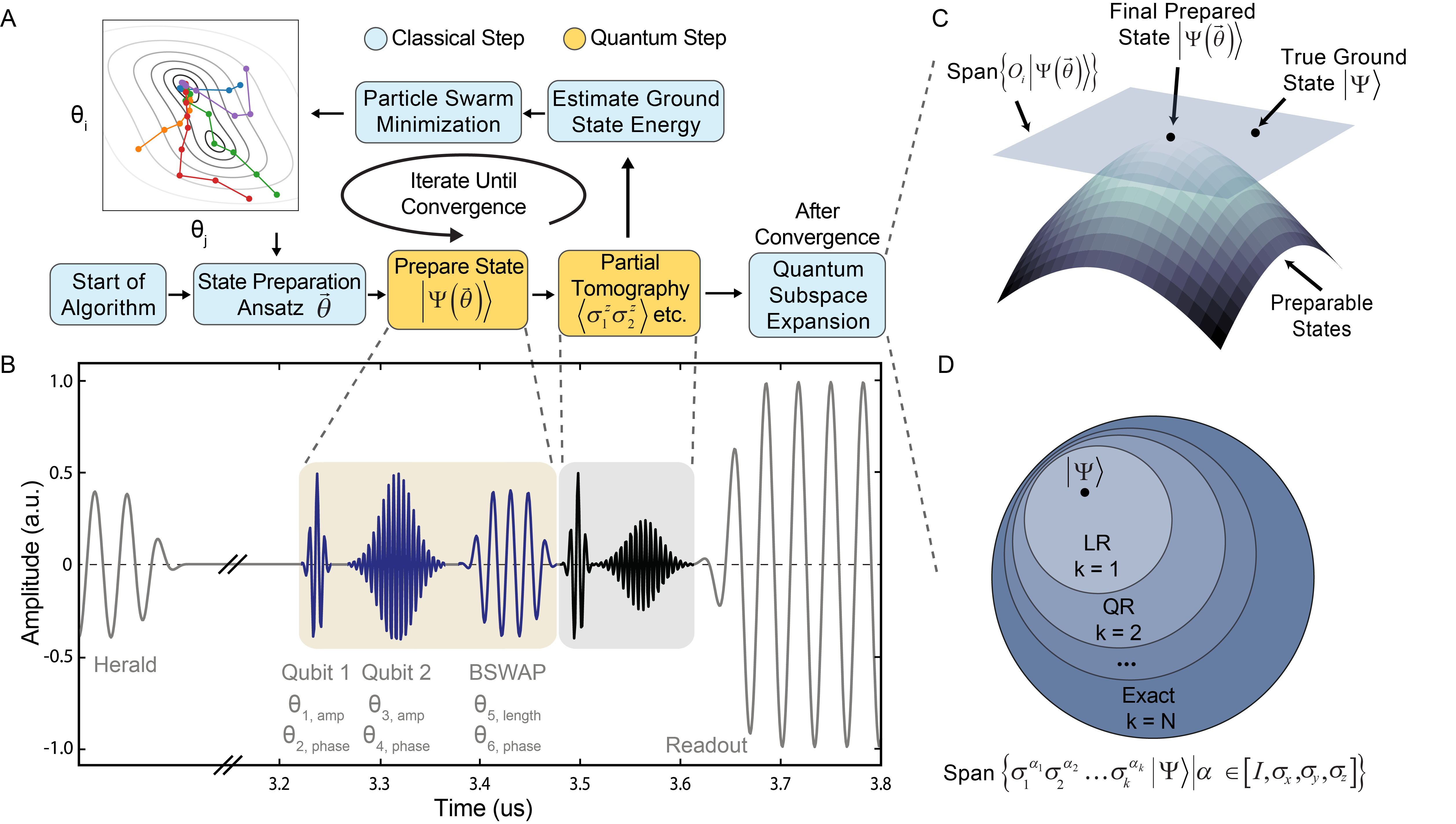}
  \caption{Description of the variational quantum eigensolver algorithm and associated quantum subspace expansion. (A) Flowchart outline of the algorithm with classical resources colored in blue and quantum resources colored in yellow. (B) Typical qubit preparation and measurement sequence consisting of a herald, single-qubit and two-qubit pulses, tomography and readout. (C) Cartoon of the QSE protocol; operators from $O_i$ are used to expand about the variational solution provided by the VQE, allowing for the mitigation of incoherent errors that otherwise render the true ground state inaccessible. (D) QSE basis hierarchy obtained from expanding about the VQE reference solution. At k = 1 one has the linear response (LR) subspace while at k = N one spans the entire subspace corresponding to the particle number of the reference state, adapted from \cite{JarrodMollie}.}
\label{fig:Fig_1}
\end{figure*}
\vspace{0.1in}
\noindent \textbf{General Approach}

\noindent The electronic structure Hamiltonian, an operator on the space of electronic wavefunctions, is first cast into a form suitable for evaluation on a quantum processor.  Specifically, the Hamiltonian is first projected onto a discrete set of molecular orbitals---we use the conventional STO-3G basis set~\cite{Hehre}, which constitutes a so-called minimal set in that it represents the minimum number of orbitals required to represent a given atomic shell. The resulting fermionic Hamiltonian $H_F$ is finally mapped onto a two-qubit Hamiltonian $H_Q$ (\emph{SI Mapping of the H$_2$ Hamiltonian to qubits}).  For the hydrogen molecule, $H_Q$ takes the form  
\begin{equation}
H_Q = g_0 + g_1 \sigma_z^1 + g_2 \sigma_z^2 + g_3 \sigma_z^1 \sigma_z^2 + g_4 \sigma_y^1 \sigma_y^2 + g_5 \sigma_x^1 \sigma_x^2,
\end{equation}
where $\sigma_k^i$ is the $k^{\mathrm{th}}$ Pauli operator on the $i^{\mathrm{th}}$ qubit, and the coefficients $\{ g_m \}$, and thus the total Hamiltonian, depend parametrically on $R$, the separation between the two hydrogen nuclei. For a given two-qubit state $|\psi\rangle$, prepared on the quantum processor, the expectation $\langle H_Q \rangle$ is evaluated through repeated measurements of Pauli correlators.  

An outline of the VQE algorithm is depicted in Fig.~\ref{fig:Fig_1}$A$ and consists of parameterizing a quantum circuit $U({\vec{\theta}})$ to prepare an ansatz wavefunction $|\psi(\vec{\theta})\rangle$, evaluating the expectation  $\langle \psi(\vec{\theta}) |H_Q| \psi(\vec{\theta}) \rangle$ term-wise using a quantum processor, and then using a classical minimization algorithm to update parameters until a minimum, $\vec{\theta}_{min}$ is found.  The quantum state $|\psi(\vec{\theta}_{min})\rangle$ then constitutes an approximation to the ground state of $H_Q$, with an estimated energy of $\langle \psi(\vec{\theta}_{min}) |H_Q| \psi(\vec{\theta}_{min}) \rangle$. 

Once the VQE algorithm has converged on an estimate of the ground state wavefunction, the quantum subspace expansion can be applied by measuring additional Pauli correlators that form an approximate matrix representation of $H_Q$ within an expanded subspace. This matrix can then be diagonalized classically to yield both low-lying excited-state energies and a refined ground-state energy estimate (Fig.~\ref{fig:Fig_1}$C$).  If the expansion is chosen such that its dimension scales polynomially with system size, this classical matrix calculation is efficient~\cite{JarrodMollie}. The effectiveness of the QSE thus requires the existence of such a subspace which captures a significant amount of the excited state support.  

We expect that molecular excited energy levels differ from the ground state primarily by excitations which promote a single electron from an occupied to an unoccupied orbital. Therefore to a good approximation, these states are linear combinations of $\{S_{1}: a_i^\dagger a_j\ket{\psi_{GS}}\}$, where $a_j$ ($a_i^\dagger$) are fermionic annihilation (creation) operators for the electronic orbitals.  While $S_1$ could serve as a subspace, a more natural choice when working with qubits involves the set of single Pauli flips $\{P_1: \sigma_\alpha^k \ket{\psi_{GS}}\ |\ \alpha \in \{x, y, z\}, k \in \{1,2\} \}$ (Fig.~\ref{fig:Fig_1}$D$), which we refer to as a linear response expansion. To calculate the matrix elements $H_{ij}$ in the $P_1$ basis, we use the quantum processor to evaluate the inner products
\begin{equation}
H_{ij} = \bra{\psi_{GS}}\sigma_i^\dagger H\sigma_j \ket{\psi_{GS}},  
\end{equation}
where $\ket{\psi_{GS}}$ is taken to be the initial approximate ground state $\ket{\psi(\vec{\theta}_{min})}$, found via the VQE routine.

Beyond providing a means of calculating molecular excited state energies, it was conjectured in ref.~\cite{JarrodMollie} that the inclusion of specific measurement operators expanding the number of states under consideration, the QSE could improve the accuracy of the initial VQE ground state estimate. While the VQE can in principle correct for the presence of coherent gate errors, the QSE was thought to additionally correct for incoherent errors, such as dephasing or amplitude damping. As discussed in the results section, we find experimental support for this conjecture. 

\vspace{0.1in}
\noindent\textbf{Experimental Methods}

\noindent \textbf{Quantum. } The quantum processor we use to evaluate expectation values consists of two superconducting qubits of the transmon variety~\cite{Koch, Paik}, one of the leading types of superconducting qubits in terms of design simplicity and coherence time.  The qubits are initialized in the joint ground state $|00\rangle$ via a heralding measurement~\cite{Johnson}. A generating circuit $U(\vec{\theta})$ is then used to prepare the desired trial wavefunction (with $\vec{\theta}$ specified by the classical hardware---see next section). 

$U(\vec{\theta})$ consists of three microwave pulses resonant with the desired qubit transition (shown in Fig.~\ref{fig:Fig_1}$B$).  First, two single-qubit rotations take place, parameterized by amplitudes ($\theta_1, \theta_3$) and phases ($\theta_2, \theta_4$).  Second, an entangling operation, known as the bSWAP gate~\cite{Poletto}, performs a rotation within the subspace spanned by $\{|00\rangle, |11\rangle\}$, parameterized by a length ($\theta_5$) and a phase ($\theta_6$). 
  
Single-qubit pulses on qubit A and B last 50 and 70 $ns$ respectively, and achieve fidelities of $\sim$99\%.  The two-qubit pulse takes up to 310 ns and approaches a fidelity of $\sim$96\%.  After state preparation via $U(\vec{\theta})$, tomographic reconstruction is used to evaluate $\left\langle H \right\rangle  = \sum\limits_{ij} {{h_{ij}}\left\langle {{\sigma _i}{\sigma _j}} \right\rangle }$.  A near-quantum-limited traveling wave parametric amplifier~\cite{macklin1, obrien1} enables high-fidelity single-shot measurement of the joint qubit state \emph{(SI Experimental Details)}.  The entire sequence, including both state preparation and measurement, comprises less than $\sim$1.5 $\mu$s, below the coherence times of the qubits: 16 $\mu$s $T_{1A}$, 13.5 $\mu$s $T^{*}_{2A}$, 12 $\mu$s $T_{1B}$, 3.5 $\mu$s $T_{2B}^{*}$.

\begin{figure}
  \includegraphics[width=\columnwidth]{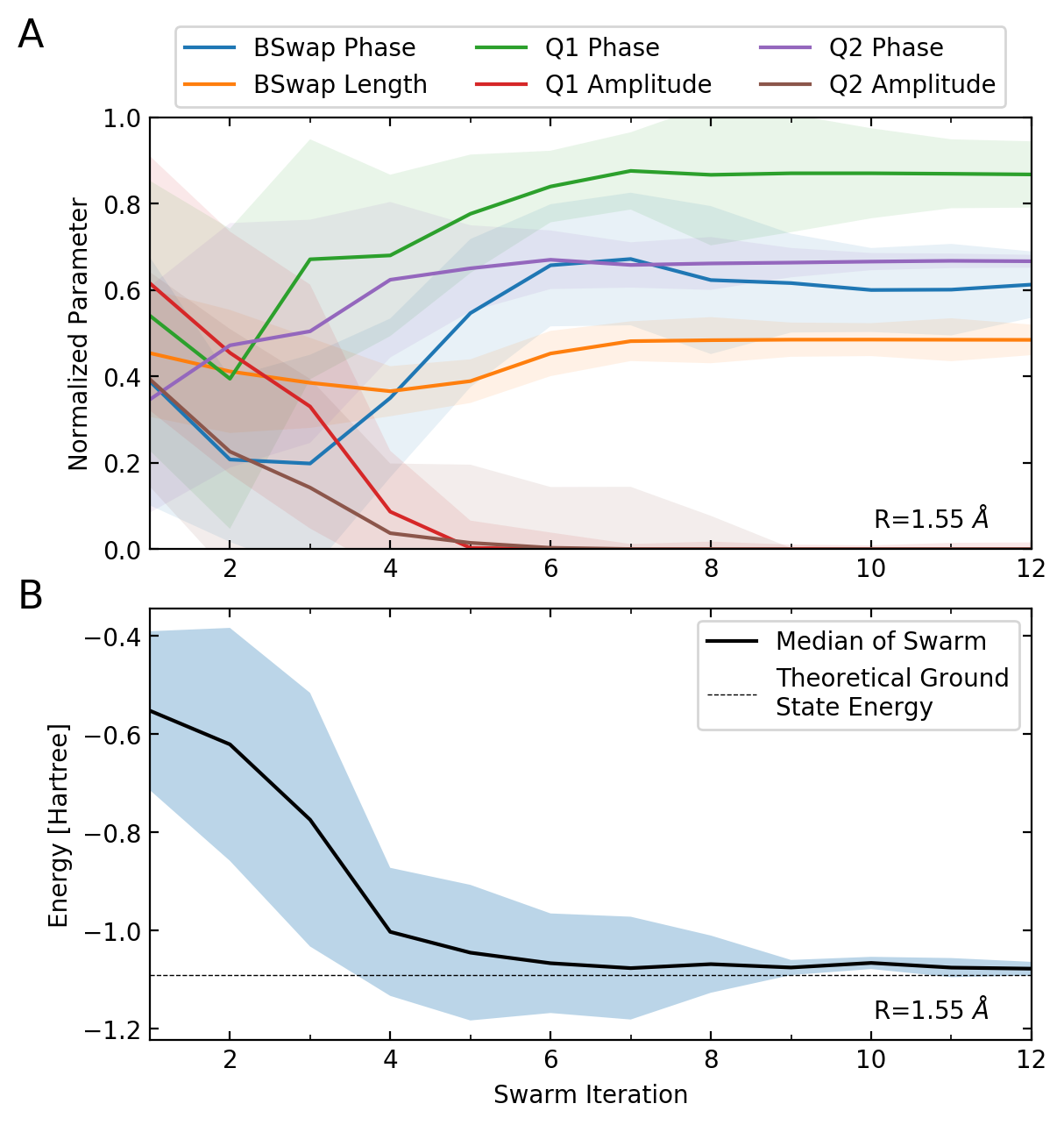}
  \caption{Control parameter convergence as a function of classical optimizer iteration. (A) Median (solid line) and standard deviation (shaded region) for all 6 normalized parameter values as a function of swarm iteration number at an internuclear bond distance of $1.55$ \r{A}. A swarm of 20 particles demonstrates convergence after approximately 12 iterations or equivalently 240 function evaluations. (B) Median energy (solid line) and standard deviation (shaded region) of swarm particles as a function of iteration number for the corresponding data in (A). Monotonic convergence of median energy towards the theoretical value is observed followed shortly thereafter by a rapid decrease in swarm energy variance.}
  \label{fig:Fig_2}
\end{figure}

\begin{figure*}
  \includegraphics[width=\textwidth]{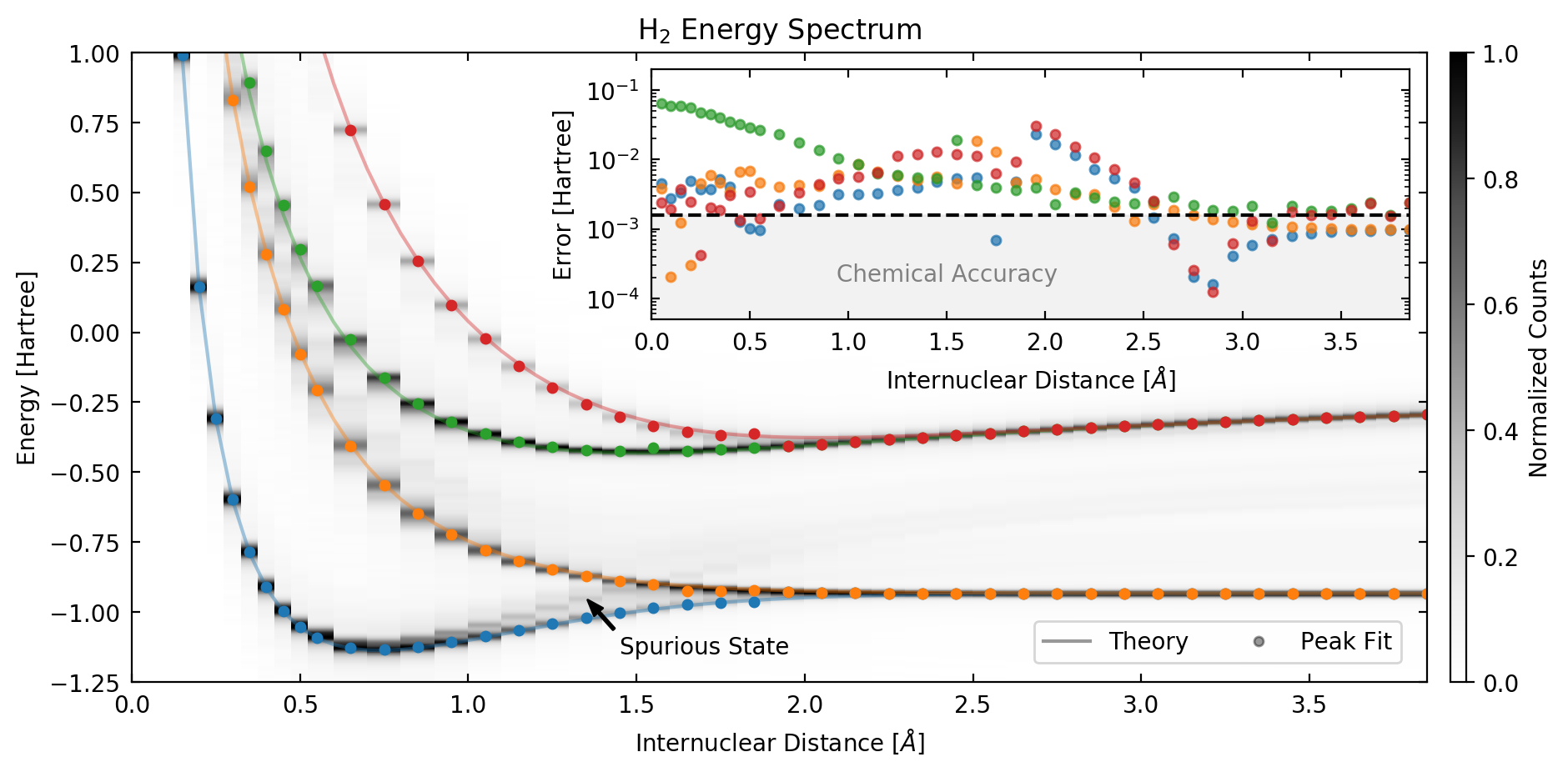}
\caption{H$_2$ energy spectrum as a function of internuclear distance. Swarm particle energies for each bond length are histogrammed after application of a linear response expansion and Gaussian filter. Energy estimates obtained by a peak finding routine are indicated by dots with theoretically predicted energy levels shown as solid lines. An unphysical spurious state emerges at internuclear distances greater than $\sim1.2$ \r{A} due to uncorrected incoherent errors. Inset shows errors in the estimated ground and excited state energies as compared to chemical accuracy ($1.6\times10^{-3}$ Ha).}
  \label{fig:Fig_3}
\end{figure*}
\vspace{0.1in}

\noindent \textbf{Classical.} With the two-qubit processor providing a means to efficiently evaluate $\langle H \rangle (\vec{\theta})$, the classical computer uses a particle-swarm optimizer (PSO) to find parameter values $\vec{\theta}_{min}$ which minimize this objective function, as shown in Fig. \ref{fig:Fig_2}$A$. The PSO approach has two properties useful for this work: it is likely to avoid getting trapped in local minima and it is more robust to noisy objective-function calls~\cite{Parsopoulos}. The optimization  treats a single point in parameter space as a particle, which has a velocity and position. A swarm of $n$ such particles $\{\ket{\psi(\vec{\theta}_{s,i})}, i\in [1,n]\}$ (with $s$ the swarm iteration number) is first randomly initialized and then at each iteration, the particles' positions are updated based on both their own energy evaluation and those of others in the swarm (\emph{SI Experimental Details}). Figure \ref{fig:Fig_2}$B$ shows how iterating through this loop allows the particles to converge on a set of control parameters that prepares the best approximation of our system's ground state $\left| {\psi ({\vec{\theta} _g})} \right\rangle$ and its associated energy.

\noindent\textbf{Results}

\noindent We apply our algorithm to the H$_2$ molecule for 45 internuclear distances between $0.05$~\r{A} and $3.85$~\r{A}.  As shown in Fig.~\ref{fig:Fig_2}$A$ for a internuclear distance of $1.55$~\r{A} and a random initialization of 20 swarm particles over $\vec{\theta}$, we observe good convergence of the control parameters within 12 swarm iterations. Each function evaluation consists of 10,000 acquisitions and lasts on the order of a minute, resulting in a total algorithm run time of approximately four hours. Experimentally optimized parameters show deviation from those that would be expected in the case of idealized gates (\emph{SI VQE and Coherent Errors}). In particular, while the experimental single-qubit gate amplitudes and two-qubit BSWAP length agree with numerical simulations, the phase of the BSWAP differs significantly, most likely due to an unaccounted for Stark shift during application of the gate. The successful convergence of the algorithm despite this mis-calibration thus provides additional proof of the protocols intrinsic ability to correct for coherent errors.

Plotting the median energy of the swarm as a function of iteration number, we observe a large initial energy error due to the random nature of the particle initialization, followed by an almost monotonic decrease towards the exact theoretical value. When calculating an estimate for a new internuclear distance, we exploit the smoothness of the parameter landscape and re-initialize the swarm particles around the minimum found in the preceding run, allowing them to vary by only 5$\%$ from their previous optimum values.  This results in subsequent runs requiring fewer resources---20 particles and 6 swarm iterations---in order to reach convergence.  Once each internuclear separation of interest has been processed, we have an initial approximation for the ground state energy function of the H$_2$ molecule. 

\begin{figure}
  \includegraphics[width=.95\columnwidth]{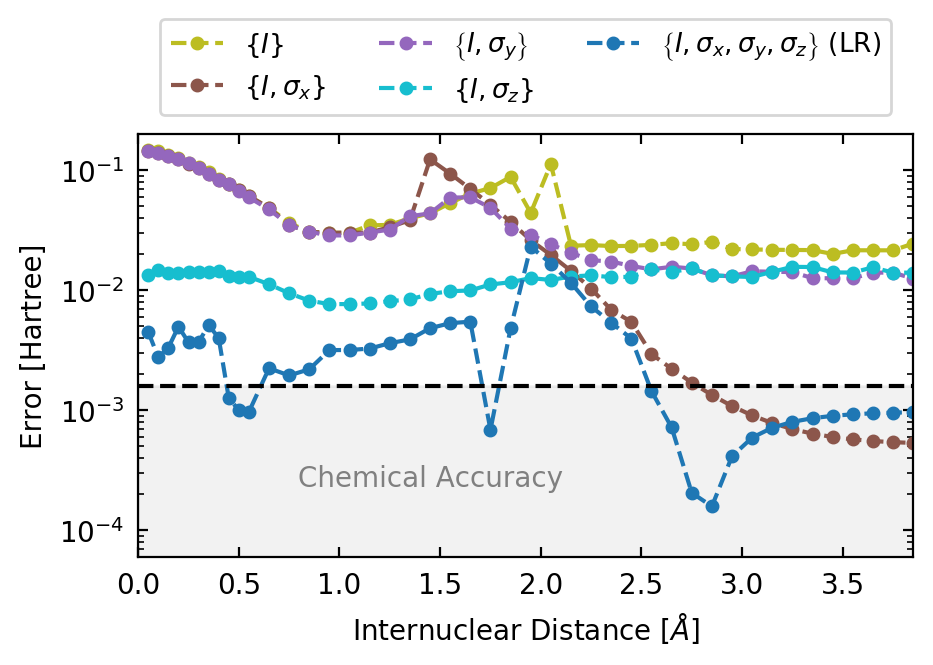}
  \caption{Comparison of errors in the ground-state energy estimate when applying the QSE protocol using  various combinations of expansion operators. The linear response expansion (dark blue dots) provides an improvement of more than an order of magnitude over the bare VQE estimate (yellow dots) for the majority of internuclear distances computed.}
  \label{fig:Fig_4}
\end{figure}

To derive excited states from this approximate ground state, we apply the linear-response QSE to each individually-reconstructed density matrix recovered during the minimization process. The results of applying this expansion are plotted in Fig.~\ref{fig:Fig_3} where data are binned with 1.5 mHa resolution before convolution with a Gaussian filter (standard deviation of 7.5 mHa). Peak-finding routines are then used to estimate the mean energies for both the corrected ground and excited states. This shows improved robustness for small numbers of swarm iterations as it is less affected by outlying particles in the swarm yet to reach the global minimum. 

Errors between experimentally predicted energies for the ground and excited states and their true values are plotted in the inset of Fig.~\ref{fig:Fig_3}. Chemical accuracy, the level required to make realistic chemical predictions, is achieved for the ground and highest excited state across a wide range of internuclear distances. Estimates of the second and third excited state energies are generally within an order of magnitude of this level.  It is interesting to note that although the ground electronic state wavefunction near equilibrium requires little entanglement to accurately represent, the same is not true of the excited states.  The QSE is able to approximate these states with only additional local measurements and efficient classical computation, without an increase in required entanglement on the quantum state of the qubits.

Figure~\ref{fig:Fig_4} shows the deviations from the theoretically expected values for the corrected ground-state energies when using different underlying measurement operators for the applied QSE. Those involving just a single Pauli operator offer sporadic improvement over the uncorrected case, with the $\sigma_z$ operator achieving best results at smaller internuclear separations while the $\sigma_x$ operator is most useful at larger ones. The complete linear-response expansion is able to mitigate incoherent errors for which that the bare VQE algorithm is unable to compensate and produces a reduction in the energy estimate error of almost two orders of magnitude over the entire range.

Note that ideally, the total number of extracted energy levels should be upper-bounded by the dimension of the Hamiltonian.  However, if the extant error channels cause the prepared VQE ground state to be sufficiently mixed (for a given set of QSE operators), it is possible to extract additional ``spurious'' energy levels.  Such a spurious state is observed as indicated in Fig.~\ref{fig:Fig_3} for internuclear distances between $\sim1.2$\r{A} and $\sim1.7$\r{A}.  In some cases, these states may be discarded on the basis of continuity of the energy as a function of internuclear distance.  Alternatively, these states can be removed by increasing the span of the QSE operators (at the cost of an increased tomographic measurement overhead). The exact conditions for the presence of a spurious state are currently being investigated.

\vspace{0.1in}
\noindent\textbf{Conclusion}

\noindent We present a novel extension of the variational quantum eigensolver that only uses a polynomial number of additional tomographic measurements to extract molecular excited states and mitigate incoherent errors on the ground state estimate. With the hydrogen molecule as a test case, we additionally confirm the intrinsic ability of the algorithm to correct for coherent gate errors when pulse properties are optimized directly. Used with classical particle swarm minimization routines well suited to high-dimensional noisy environments, these techniques yield ground- and excited-state energy estimates with near-chemical accuracy. Our results highlight the potential of QSE to significantly reduce the need for more advanced error correction techniques, thereby facilitating practical applications of near-term quantum hardware.

\vspace{0.1in}
\noindent\textbf{Acknowledgements}

\noindent Financial support for ongoing quantum circuit development was provided by ARO/LPS under grant W911NF-15-1-0496 and for the current experiment by the Director, Office of Science, Office of Advanced Scientific Computing Research, of the U.S. Department of Energy under Contract No. DE-AC02-05CH11231.

\bibliography{Main}
\bibliographystyle{pnas-new}

\pagebreak

\widetext
\begin{center}
\textbf{\large Supplemental Materials: Robust determination of molecular spectra on a quantum processor}
\end{center}
%%%%%%%%%% Merge with supplemental materials %%%%%%%%%%
%%%%%%%%%% Prefix a "S" to all equations, figures, tables and reset the counter %%%%%%%%%%
\setcounter{equation}{0}
\setcounter{figure}{0}
\setcounter{table}{0}
\setcounter{page}{1}
\makeatletter
\renewcommand{\theequation}{S\arabic{equation}}
\renewcommand{\thefigure}{S\arabic{figure}}

\noindent\textbf{SI: Experimental Details}

\begin{figure*}
  \includegraphics[width=.95\textwidth]{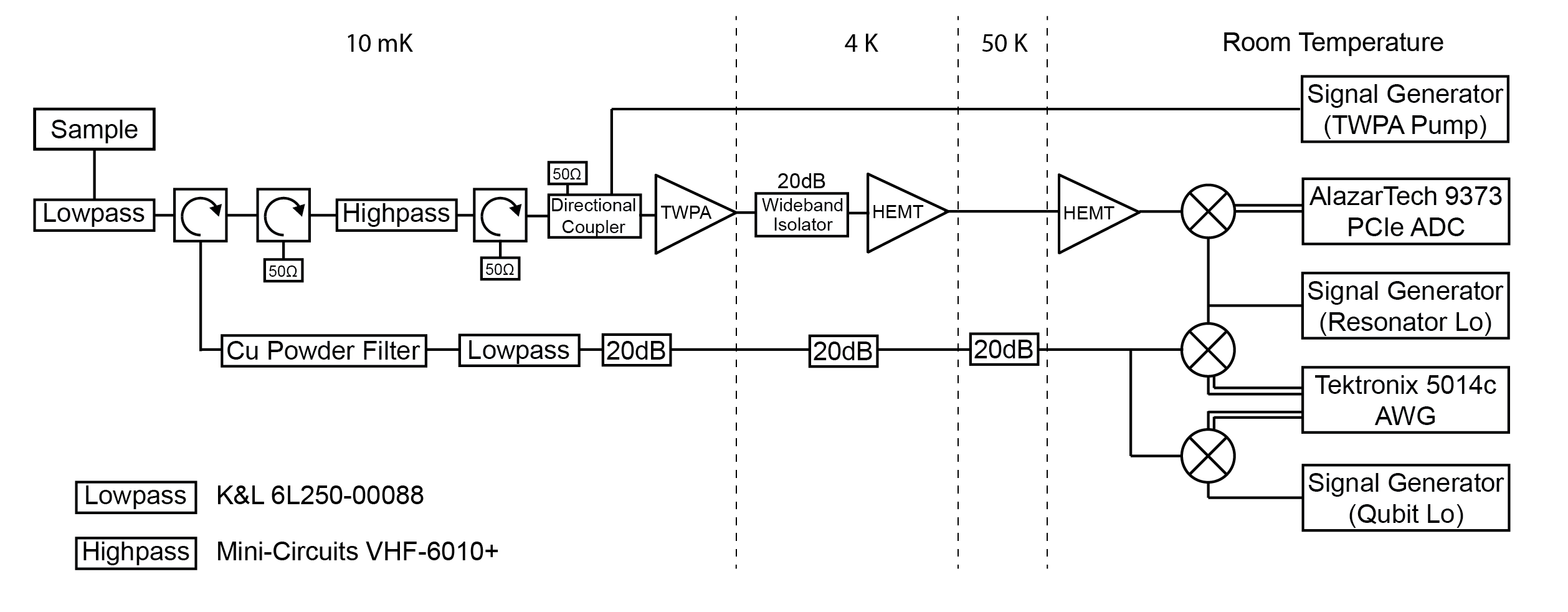}
  \caption{Schematic of the measurement setup used in the experiment.  At the 10 mK stage of the dilution refrigerator, the sample is connected in a reflection geometry to the 50 ohm environment from which it receives qubit/cavity pulses.  These pulses are generated at room temperature by the electronics shown on the right of the figure, and pass through several stages of attenuation on the way to the sample.  To enable high-fidelity measurement of the qubit state, a near-quantum-limited Traveling Wave Parametric Amplifier (TWPA)~\cite{macklin1} amplifies the signal after reflection off the cavity.  Further amplification is provided by a HEMT amplifier at the 4 K state of the dilution refrigerator, after which the signal is amplified at room temperature, downconverted to a heterodyne frequency of 11 MHz, and digitized by an AlazarTech 9373 ADC.  From this data, the qubit state is determined in software.}
  \label{fig:Supp_Fig_1}
\end{figure*}

\noindent Here we provide details of the quantum processor used in the experiment.  The device consists of two superconducting transmon qubits~\cite{Koch} on a single silicon chip, mounted in and coupled to a three-dimensional copper cavity~\cite{Paik}.  Each transmon consists of a capacitor shunted by a nonlinear inductance; in our device one of the qubits uses a single Josephson junction as the nonlinear inductance, with a fixed frequency of 3.788 GHz, while the other uses a SQUID loop, allowing for tuning the frequency (via an external magnetic field) from a zero-flux value of roughly 5 GHz to the working frequency of 4.111 GHz.  The copper cavity exhibits a resonant frequency of 7.122 GHz with a loaded linewidth $\kappa_{ext} \approx 8~ \mathrm{MHz}$, set primarily by the coupling (in a reflection geometry) to a 50-ohm environment (see Fig. \ref{fig:Supp_Fig_1}).  The cavity is mounted at the 10 mK stage of a dilution refrigerator.  

We detect the state of the qubits by using a heterodyne measurement (at heterodyne frequency $11 ~\mathrm{MHz}$) of the resonant cavity frequency, exploiting the dispersive shift between the qubits and cavity.  Because the two-qubits are coupled to a single cavity, the dispersive shift is roughly equal in magnitude for both qubits, and thus distinguishing the states $\ket{01}$ and $\ket{10}$ with single-shot fidelity is impossible.  However, our measurement is able to distinguish the joint two-qubit ground state $\ket{00}$ from all other computational states, which is sufficient for reconstructing the Pauli correlators necessary for evaluating the expectation value $\langle H \rangle$.  To evaluate $\langle H \rangle$ we first reconstruct the two-qubit density matrix using a set of 32 tomographic measurements (see \cite{Chow} for details), then calculate the necessary correlators given the density matrix.  In future implementations of VQE on larger quantum systems, full tomography will be impossible (due to an exponential scaling of the number of required measurements).  Instead, only the necessary correlators will be directly measured.  For this reason, our reconstruction of the two-qubit density matrix from the tomographic measurements did not use any method such as maximum-likelihood estimation which enforces physicality (positivity and trace-normalization) on the result.  

For the classical optimization routine, we used the Pyswarm package, a Python implementation of particle-swarm optimization.

\vspace{0.1in}
\noindent\textbf{SI: Mapping of the H$_2$ Hamiltonian to Qubits}

\noindent The H$_2$ Hamiltonian was determined by calculation of the 
electronic integrals in the standard Gaussian STO-3G basis~\cite{Hehre}.  For a hydrogen atom, this basis set consists of the single $1s$ orbital.  
After determination of the one- and
two-electron integrals, the Hamiltonian was projected into a particle conserving
manifold defined by a determinantal configuration interaction expansion that does
not flip spins.  That is, within a molecular orbital basis we define spatial orbitals
$1$ and $2$ with possible spins $\alpha$ and $\beta$ such that the 4 spin-orbitals
in the system can be populated by the second quantized operators $a_{1\alpha}^\dagger, a_{1 \beta}^\dagger, a_{2 \alpha}^\dagger, a_{2 \beta}^\dagger$.  Beginning from the reference state defined by $a_{1\alpha}^\dagger a_{1 \beta}^\dagger \ket{vac}$, where $\ket{vac}$ is the
Fermi vacuum state.  This generates the
following four basis states that we map to computational basis 
states explicitly as follows
\begin{align}
a_{1\alpha}^\dagger a_{1\beta}^\dagger \ket{vac} &\rightarrow \ket{00} \\
a_{1\alpha}^\dagger a_{2\beta}^\dagger \ket{vac} &\rightarrow \ket{01} \notag \\
a_{2\alpha}^\dagger a_{1\beta}^\dagger \ket{vac} &\rightarrow \ket{10} \notag \\
a_{2\alpha}^\dagger a_{2\beta}^\dagger \ket{vac} &\rightarrow \ket{11}. \notag
\end{align}
We note 
that such a reduction to 2-qubits or fewer can also be achieved either through the 
Bravyi-Kitaev transformation, as noted by O'Malley et al.~\cite{Omalley} or through 
alternative symmetry enforcing transformations as introduced by 
Bravyi et al~\cite{Bravyi}. 
The Hamiltonian in this space was then expressed in the basis of Pauli operators to
yield a Hamiltonian of the form:
\begin{align}
H_Q(R) = \sum_{ij}^{\alpha \beta} g_{ij} (R) \sigma^i_\alpha \sigma^j_\beta
\end{align}
for each nuclear configuration R, where $\sigma^i_\alpha$ is a Pauli operator 
acting on qubit $i$ from $\sigma^i_\alpha \in \{I^i, \sigma^i_x, \sigma^i_y, \sigma^i_z \}$.  Due to additional spatial, spin, and time-reversal symmetry in
the molecular Hamiltonian, many of the coefficients are $0$ for all nuclear
configurations $R$ and all are real-valued. 
As a result, the Hamiltonian may be more compactly expressed as
\begin{align}
H_Q(R) &= g_0(R) + g_1(R) \sigma_z^1 + g_2(R) \sigma_z^2  \notag \\
& + g_3(R) \sigma_z^1 \sigma_z^2
+ g_4(R) \sigma_y^1 \sigma_y^2 + g_5(R) \sigma_x^1 \sigma_x^2.
\end{align}
The exact coefficients used in this work are given in Table \ref{tab:Hamiltonian} of this SI.  

\vspace{0.1in}
\noindent\textbf{SI: QSE and Choice of Expansion Operators}

\begin{figure*}
  \includegraphics[width=.95\textwidth]{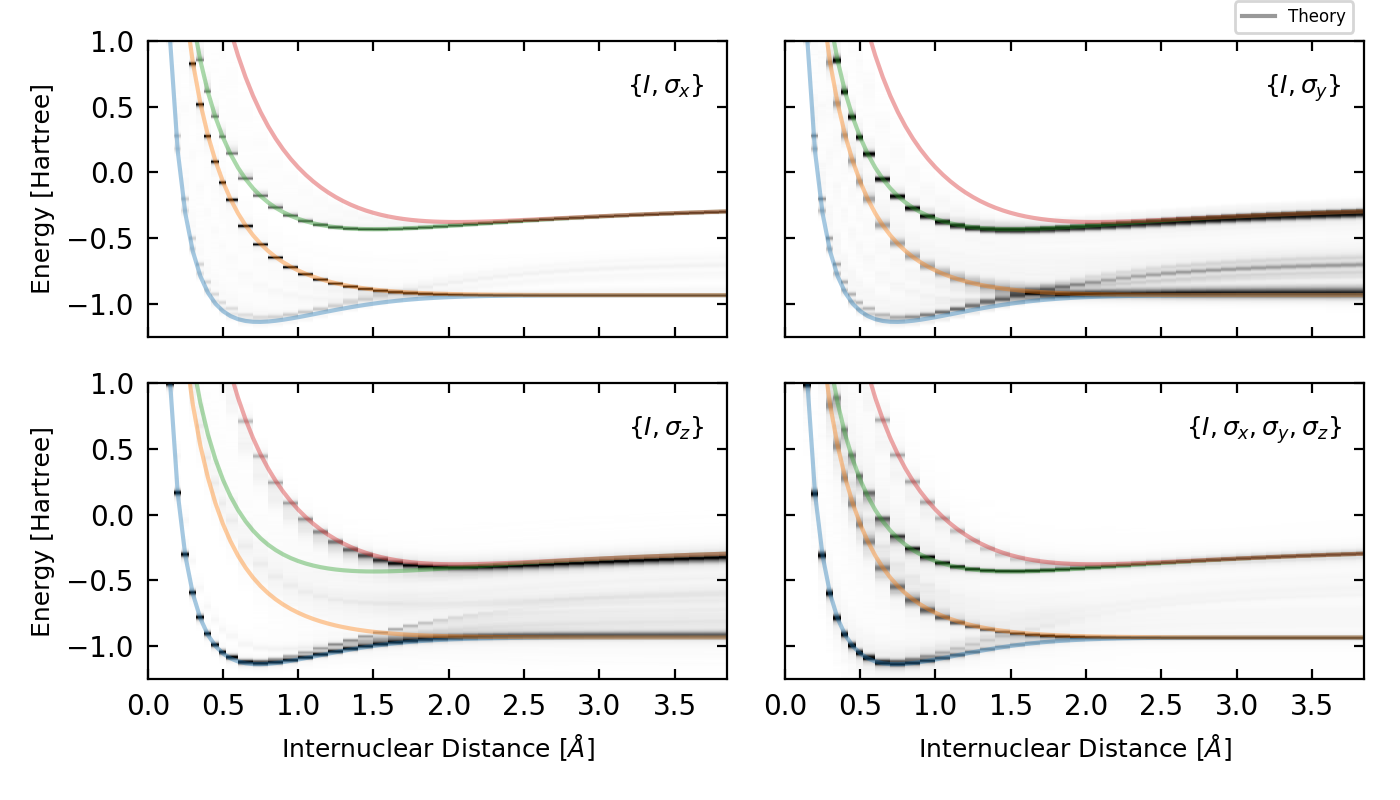}
 \caption{QSE protocol applied with different choices of measurement operators. Using only the identity and $\sigma_y$, $\sigma_x$, or $\sigma_z$ results in errors in the calculated excited energies while using a full linear response expansion successfully resolves the entire spectrum.}
  \label{fig:Supp_Fig_2}
\end{figure*}

\noindent The choice of operators which act on the ground-state density matrix to form the expanded subspace influences which excited can be extracted.  This we show in figure~\ref{fig:Supp_Fig_2}.  Using only the identity and single Pauli operators (on each qubit) results in only a partial resolution of the low-lying excited states, while the full linear-response is able to resolve the entire spectrum.

\vspace{0.1in}
\noindent\textbf{SI: QSE with Errors}

\noindent The quantum subspace expansion (QSE), works by resolving the action of an operator $H$ within a subspace defined by a set of operator $\{O_i\}$, such as the single fermion excitation operators $S_1$ or $k^{\mathrm{th}}$ order Pauli operators $P_k$ defined in the text.  This is done by measuring the matrix elements coupling the states generated by these operators $H_{ij} = \bra{\psi_0}O_i^\dagger H O_j \ket{\psi_0}$ as well as the corresponding identity operator in this space, also known as the overlap matrix, $S_{ij} = \bra{\psi_0}O_i^\dagger O_j \ket{\psi_0}$.  The action within this subspace is then used to provide increasingly accurate approximations (as a function of the subspace size) by solving the generalized eigenvalue problem $HC=SCE$ with the matrices $H$ and $S$ for the eigenvectors $C$ and diagonal matrix of eigenvalues $E$.

Defining the density matrix for the pure state $\ket{\psi_0}$ as $\rho_0=\ket{\psi_0}\bra{\psi_0}$, it is easy to see these matrix elements are equivalent are equivalent to $H_{ij} = \text{Tr}[O_i^\dagger H O_j \rho_0]$ and $S_{ij} = \text{Tr}[O_i^\dagger O_j \rho_0]$.  This formulation naturally generalizes to mixed states $\rho$ with rank $>1$, which are the case in essentially any real system with incoherent errors, and gives a clear prescription for the measurement of the matrix elements.  However, in this notation it is less clear how the measurement correspond to action within a subspace and what this means in the case for mixed states $\rho$ with rank $> 1$.  To clarify these situations, we may alternatively use the vectorization of the density matrix to re-express these matrix elements.

We denote the row-major vectorization~\cite{gilchrist2009vectorization} of a matrix $\rho$ as $\ket{\rho}\rangle$.  In this notation, we have that 
\begin{align}
H_{ij} &= \text{Tr}[O_i^\dagger H O_j \rho] \notag \\
&= \langle \langle O_i \ket{H O_j \rho} \rangle \notag \\
&= \langle \langle O_i | H \otimes \rho^T \ket{O_j} \rangle \\
S_{ij} &= \langle \langle O_i | I \otimes \rho^T \ket{O_j} \rangle.
\end{align}
This construction clarifies a number of the mathematical properties, including the hermiticity of the matrices and their dimensionality.  In the case of a pure state, the operator $\rho^T$ has a single non-zero eigenvalue, and the maximum non-trivial dimension of the space, which is determined by the trace of the identity operator $S$ using normalized operators $\{O_i\}$, is that of the Hamiltonian.  

It is important to consider in more detail when the rank of $\rho$ is $>1$.  In these cases, the dimension of the space is potentially greater than the dimension of the original Hamiltonian.  The easiest way to see this is to consider the case of the maximally mixed state $\rho = \frac{1}{d}I$, where $d$ is the dimension.  In this case, the dimension determined by the identity is the square of that of the original Hamiltonian, which this construction makes clear is the maximal dimension of this problem.  Moreover, by properties of the standard tensor product, it is easy to verify that the eigenvalues of $H \otimes I$ are the eigenvalues of $H$, but $d-$fold degenerate.  We note that the factor of $1/d$ is treated by its appearance in the metric matrix $S$ in the generalized eigenvalue problem.

Thus, if one measures a linearly independent, complete set of operators $\{O_i\}$ on the totally mixed state, the resulting eigenvalues will be the spectrum of $H$ with $d-$fold degeneracies.  These additional states represent the different possible expansions from components $\ket{\psi_i}\bra{\psi_i}$ with $\rho = \sum_i \ket{\psi_i}\bra{\psi_i}$ that allowed one to prepare the eigenstates using $\{O_i\}$.  

Consider, however, if the resolution of the operator $\rho$ is incomplete with the given measurements $\{O_i\}$, then one may have difficulty distinguishing the eigenstate approximations generated from different pure states.  This leads to the so-called ``spurious states'' observed experimentally in this work, which are extra predicted eigenvalues that do not coincide with the eigenvalues of $H$.

To clarify these effects, consider the following Hamiltonian
\begin{align}
H = \sigma_z^1 + \sigma_z^2 + \sigma_x^1 \sigma_x^2
\end{align}
which has a ground state given by $\rho_0=\alpha \ket{00} + \beta \ket{11}$.  In the case of $\rho_0$, it is clear that the maximum dimension of this problem is $4$ with any set of measurement.  

Now considering the mixed state generated by a Pauli-$X$ channel that occurs with probability $p \neq 0,1$, $\rho = (1-p) \rho_0 + p \sigma_x^1 \rho_0 \sigma_x^1$.  In this case, the operator $\rho$ has as non-trivial eigenvalues $p$ and $(1-p)$.  As an example we choose $p=\frac{1}{2}$ such that it has a degenerate non-trivial spectrum of $\frac{1}{2}$ and $\frac{1}{2}$. In the case one measures a complete set of operators $\{O_i\}$, one then finds the eigenvalues of $H$ with a degeneracy of $2$ in each case.  If we consider only the error in the estimate of the ground state energy, one finds that the operator set $\{I, \sigma_z^1 \sigma_z^2\}$ is sufficient to correct it exactly (if applied to the state resulting via acting with the error channel on the ideal ground state, i.e. without minimization on this value).  The exact condition for the set of operators that correct errors in the ground state for a given $H$ and error channel and their relation to traditional theories of quantum error correction is an open problem, currently the subject of ongoing research. We conjecture here based on numerical observations that conditions are related to the ability to construct operators within Span$(\{O_i\})$ that both commute with $H$ but not with the error channel $E$.

To study the case of spurious states, suppose one measures a set of operators with dimension greater than the dimension of the Hamiltonian but not sufficient to resolve $\rho$.  In this case, one such set is $\{I, \sigma_x^1, \sigma_y^1, \sigma_x^2, \sigma_z^2, \sigma_x^1 \sigma_x^2, \sigma_x^1 \sigma_z^2, \sigma_y^1 \sigma_x^2, \sigma_y^1 \sigma_z^2\}$.  If one measures the Hamiltonian and overlap on state $\rho$ with this basis, one sees examples of the observed behaviors.  

First, the non-trivial dimension of the problem is $7$, which would be an experimental signature that the measured state is not a pure state but also not the totally mixed state.  Second, the eigenspectrum contains the exact spectrum of $H$, but is not degenerate. Rather it contains $3$ erroneous eigenvalues that correspond to the spurious states we define above.  Thus the total spectrum is formed from a combination of an exact expansion from one state and a poor expansion from another.  If one continues to add operators, the spurious values disappear, replaced by degeneracies in the spectrum on the exact values.  It is interesting to note that if one chooses operators capable of correcting these errors, a smaller set such as $\{I, \sigma_x^1, \sigma_z^1, \sigma_z^2, \sigma_x^1 \sigma_z^2, \sigma_z^1 \sigma_z^2\}$ produces the exact spectrum with degeneracies only on the $2$nd and $3$rd eigenvalues with no spurious states. 

\vspace{0.1in}
\noindent\textbf{SI: VQE and Coherent Errors}

\begin{figure}
  \includegraphics[width=0.5\columnwidth]{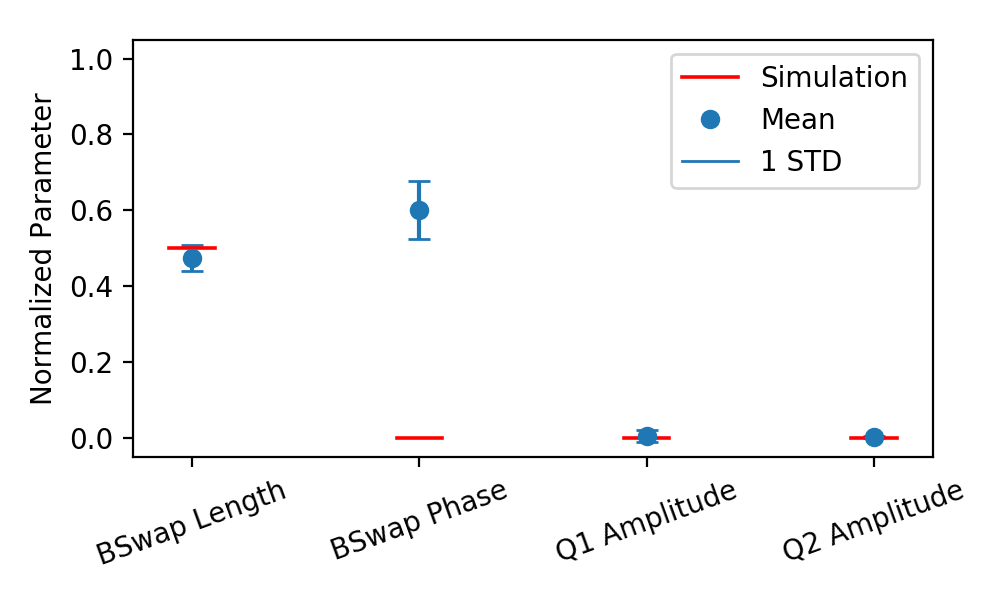}
  \caption{Final converged parameter values for 20 swarm particles at internuclear distance of 1.55 $\AA$ with results of a numerical simulation shown in red. Single qubit phases are not included as the amplitudes have converged to 0, rendering them arbitrary.}
  \label{fig:Supp_Fig_4}
\end{figure}

\noindent The VQE is expected to have an intrinsic ability to correct for coherent gate errors (such as under or over rotations) due to the direct parameterization of the microwave pulse amplitudes/lengths and phases.  As a signature of this ability, we plot in Fig. \ref{fig:Supp_Fig_4} the optimal parameters found by the VQE algorithm (for an internuclear distance of 1.55 $\AA$) and compare them to the parameters expected from our initial simulations.  The amplitudes of the single-qubit rotations converge to nearly zero, as expected from simulations.  The length of the bSWAP gate, which is finite so as to create entanglement between the two qubits, also agrees with simulation.  However, the phase of the bSWAP drive differs significantly from the expected value, namely zero.   To say this another way, at this internuclear separation, the theoretically-expected ground state wavefunction is a superposition of the states $|00\rangle$ and $|11\rangle$ with equal phases, yet the experimentally prepared ground state clearly exhibits a phase difference in the amplitudes of the $|00\rangle$ and $|11\rangle$ states.  This discrepancy is likely due to an uncalibrated Stark shift during the bSWAP gate which is corrected for automatically by the classical minimization routine, which has no knowledge of the true bSWAP unitary transformation. Phases of the single qubit drives are not included on the figure, as the single qubit amplitudes have converged to zero, which renders the phase meaningless.

\vspace{0.1in}
\noindent\textbf{SI: QSE beyond Linear Response}

\begin{figure}
  \includegraphics[width=0.5\columnwidth]{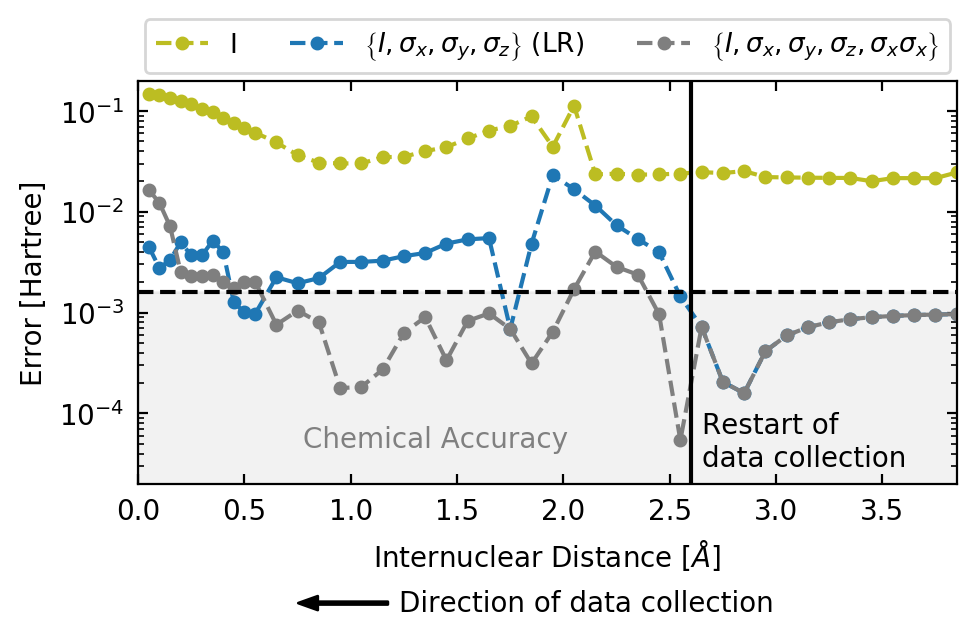}
  \caption{Starting from a large initial internuclear distance, the purely linear response expansion (blue) achieves chemical accuracy in the calculation of the ground state energy until a restart of the data collection run due to technical reasons (vertical black line). From this point onwards the accuracy in linear response estimate is degraded, most likely due to calibration errors in our tomographic reconstructions. Overcoming this calibration drift can be achieved by including additional two-qubit correlators such as $\sigma_x\sigma_x$ in the measurement span (gray).}
  \label{fig:Supp_Fig_3}
\end{figure}

\noindent In this two-qubit example of the QSE, it is straightforward to go beyond the linear-response subspace and include additional measurement operators (see Fig. \ref{fig:Supp_Fig_3}), as a demonstration that further error mitigation is possible.  The dataset shown in the figure was taken in two separate runs: the first, for internuclear separations greater than 2.6 Angstroms, and the second for separations lower than that value.  Technical reasons necessitated a restart of the data collection at that value.  In this dataset, the bare VQE ground state error is more than an order of magnitude above the threshold for chemical accuracy.  In the initial data run (for internuclear separations greater than 2.6 Angstroms), the linear-response correction is able to bring this error down below the chemical-accuracy threshold.  However, after the restart, the linear-response is no longer able to get below chemical accuracy.  The likely reason for this is a drift in the gates which effected the tomographic reconstructions.  But even though the linear response fails to fall below chemical accuracy, such accuracy can still be achieved over a large range of the separations by including additional operators in the QSE.  In the figure, specifically, we show how the addition of the operator $\sigma_x^1\sigma_x^2$ dramatically improves the accuracy of the ground-state estimate.

\begin{center} 
\begin{table}
\begin{tabular}{|c|c|c|c|c|c|}
\hline
R ($\AA$) & $I$ & $\sigma_z^1$ & $\sigma_x^1 \sigma_x^2$ & $\sigma_z^2$ & $\sigma_z^1 \sigma_z^2$ \\
\hline
0.05 & 1.00777E+01& -1.05533E+00& 1.55708E-01& -1.05533E+00& 1.39333E-02\\
0.10 & 4.75665E+00& -1.02731E+00& 1.56170E-01& -1.02731E+00& 1.38667E-02\\
0.15 & 2.94817E+00& -9.84234E-01& 1.56930E-01& -9.84234E-01& 1.37610E-02\\
0.20 & 2.01153E+00& -9.30489E-01& 1.57973E-01& -9.30489E-01& 1.36238E-02\\
0.25 & 1.42283E+00& -8.70646E-01& 1.59277E-01& -8.70646E-01& 1.34635E-02\\
0.30 & 1.01018E+00& -8.08649E-01& 1.60818E-01& -8.08649E-01& 1.32880E-02\\
0.35 & 7.01273E-01& -7.47416E-01& 1.62573E-01& -7.47416E-01& 1.31036E-02\\
0.40 & 4.60364E-01& -6.88819E-01& 1.64515E-01& -6.88819E-01& 1.29140E-02\\
0.45 & 2.67547E-01& -6.33890E-01& 1.66621E-01& -6.33890E-01& 1.27192E-02\\
0.50 & 1.10647E-01& -5.83080E-01& 1.68870E-01& -5.83080E-01& 1.25165E-02\\
0.55 & -1.83734E-02& -5.36489E-01& 1.71244E-01& -5.36489E-01& 1.23003E-02\\
0.65 & -2.13932E-01& -4.55433E-01& 1.76318E-01& -4.55433E-01& 1.18019E-02\\
0.75 & -3.49833E-01& -3.88748E-01& 1.81771E-01& -3.88748E-01& 1.11772E-02\\
0.85 & -4.45424E-01& -3.33747E-01& 1.87562E-01& -3.33747E-01& 1.04061E-02\\
0.95 & -5.13548E-01& -2.87796E-01& 1.93650E-01& -2.87796E-01& 9.50345E-03\\
1.05 & -5.62600E-01& -2.48783E-01& 1.99984E-01& -2.48783E-01& 8.50998E-03\\
1.15 & -5.97973E-01& -2.15234E-01& 2.06495E-01& -2.15234E-01& 7.47722E-03\\
1.25 & -6.23223E-01& -1.86173E-01& 2.13102E-01& -1.86173E-01& 6.45563E-03\\
1.35 & -6.40837E-01& -1.60926E-01& 2.19727E-01& -1.60926E-01& 5.48623E-03\\
1.45 & -6.52661E-01& -1.38977E-01& 2.26294E-01& -1.38977E-01& 4.59760E-03\\
1.55 & -6.60117E-01& -1.19894E-01& 2.32740E-01& -1.19894E-01& 3.80558E-03\\
1.65 & -6.64309E-01& -1.03305E-01& 2.39014E-01& -1.03305E-01& 3.11545E-03\\
1.75 & -6.66092E-01& -8.88906E-02& 2.45075E-01& -8.88906E-02& 2.52480E-03\\
1.85 & -6.66126E-01& -7.63712E-02& 2.50896E-01& -7.63712E-02& 2.02647E-03\\
1.95 & -6.64916E-01& -6.55065E-02& 2.56458E-01& -6.55065E-02& 1.61100E-03\\
2.05 & -6.62844E-01& -5.60866E-02& 2.61750E-01& -5.60866E-02& 1.26812E-03\\
2.15 & -6.60199E-01& -4.79275E-02& 2.66768E-01& -4.79275E-02& 9.88000E-04\\
2.25 & -6.57196E-01& -4.08672E-02& 2.71512E-01& -4.08672E-02& 7.61425E-04\\
2.35 & -6.53992E-01& -3.47636E-02& 2.75986E-01& -3.47636E-02& 5.80225E-04\\
2.45 & -6.50702E-01& -2.94924E-02& 2.80199E-01& -2.94924E-02& 4.36875E-04\\
2.55 & -6.47408E-01& -2.49459E-02& 2.84160E-01& -2.49459E-02& 3.25025E-04\\
2.65 & -6.44165E-01& -2.10309E-02& 2.87881E-01& -2.10309E-02& 2.38800E-04\\
2.75 & -6.41011E-01& -1.76672E-02& 2.91376E-01& -1.76672E-02& 1.73300E-04\\
2.85 & -6.37971E-01& -1.47853E-02& 2.94658E-01& -1.47853E-02& 1.24200E-04\\
2.95 & -6.35058E-01& -1.23246E-02& 2.97741E-01& -1.23246E-02& 8.78750E-05\\
3.05 & -6.32279E-01& -1.02318E-02& 3.00638E-01& -1.02317E-02& 6.14500E-05\\
3.15 & -6.29635E-01& -8.45958E-03& 3.03362E-01& -8.45958E-03& 4.24250E-05\\
3.25 & -6.27126E-01& -6.96585E-03& 3.05927E-01& -6.96585E-03& 2.89500E-05\\
3.35 & -6.24746E-01& -5.71280E-03& 3.08344E-01& -5.71280E-03& 1.95500E-05\\
3.45 & -6.22491E-01& -4.66670E-03& 3.10625E-01& -4.66670E-03& 1.30500E-05\\
3.55 & -6.20353E-01& -3.79743E-03& 3.12780E-01& -3.79743E-03& 8.57500E-06\\
3.65 & -6.18325E-01& -3.07840E-03& 3.14819E-01& -3.07840E-03& 5.60000E-06\\
3.75 & -6.16401E-01& -2.48625E-03& 3.16750E-01& -2.48625E-03& 3.60000E-06\\
3.85 & -6.14575E-01& -2.00063E-03& 3.18581E-01& -2.00062E-03& 2.27500E-06\\
3.95 & -6.12839E-01& -1.60393E-03& 3.20320E-01& -1.60392E-03& 1.42500E-06\\
\hline

\end{tabular}
\caption{Coefficients defining the Hamiltonian $H_Q$ as a function of the internuclear distance $R$.  The electronic integrals were calculated in the STO-3G basis and columns with entries that were $0$ by symmetry have been omitted.}
\label{tab:Hamiltonian}
\end{table}
\end{center}

\end{document}